\documentclass[conference]{IEEEtran}
\IEEEoverridecommandlockouts

\usepackage{cite}
\usepackage{amsmath,amssymb,amsfonts}

\usepackage{graphicx}
\usepackage{textcomp}
\usepackage{booktabs}
\usepackage{threeparttable}
\usepackage{multirow}
\usepackage{url}
\usepackage{hyperref}
\usepackage{algorithm}
\usepackage{algpseudocode}
\usepackage[most]{tcolorbox}
\usepackage{xcolor}
\usepackage{subcaption}
\usepackage{tabularx}
\usepackage{makecell}
\usepackage{ragged2e}
\usepackage{array}
\usepackage{xcolor}
\usepackage{enumitem}
\usepackage[most]{tcolorbox}
\definecolor{roundbg}{RGB}{245,245,250}
\newtcolorbox{roundbox}[1]{
  colback=roundbg, colframe=black!40, 
  fonttitle=\bfseries\small, title={#1},
  boxrule=0.4pt, arc=2pt, left=4pt, right=4pt, 
  top=2pt, bottom=2pt, fontupper=\small
}
\usepackage[most]{tcolorbox}

\newtcolorbox{rqbox}[1]{
    breakable,
    colback=gray!10,      
    colframe=gray!50,     
    arc=2pt,              
    boxrule=0.8pt,        
    left=5pt, right=5pt, top=5pt, bottom=5pt, 
    fonttitle=\bfseries,  
    title=#1              
}
\usepackage{siunitx}
\newtcolorbox{agentdialogue}[2][]{
    enhanced,
    before skip=10pt,after skip=10pt,
    colback=white,
    colframe=#2!70!black, 
    coltitle=white,
    fonttitle=\bfseries\sffamily,
    title={#1},
    attach boxed title to top left={xshift=3mm, yshift=-2mm},
    boxed title style={colback=#2!70!black},
    sharp corners,
    drop shadow,
    left=2mm, right=2mm, top=3mm, bottom=2mm
}

\def\BibTeX{{\rm B\kern-.05em{\sc i\kern-.025em b}\kern-.08em
    T\kern-.1667em\lower.7ex\hbox{E}\kern-.125emX}}

\usepackage{tikz}
\usetikzlibrary{shapes.geometric, arrows.meta, positioning, fit, calc}
\usepackage{xcolor}

\definecolor{phase1color}{RGB}{230,240,255}  
\definecolor{phase2color}{RGB}{255,240,230}  
\definecolor{phase3color}{RGB}{240,255,230}  
\definecolor{phase4color}{RGB}{255,245,240}  
\definecolor{phase5color}{RGB}{245,240,255}  
\definecolor{inputcolor}{RGB}{250,250,250}   

\begin{document}

\title{QUARE: Quality-Aware Requirements Analysis through Multi-Agent Dialectical Negotiation}

\author{
\IEEEauthorblockN{Haowei Cheng}
\IEEEauthorblockA{\textit{Waseda University} \\
Tokyo, Japan \\
haowei.cheng@fuji.waseda.jp}
\and
\IEEEauthorblockN{Milhan Kim}
\IEEEauthorblockA{\textit{Waseda University} \\
Tokyo, Japan\\
milhan.kim@ruri.waseda.jp}

\and
\IEEEauthorblockN{Foutse Khomh}
\IEEEauthorblockA{\textit{Polytechnique Montréal} \\
 Montréal, Canada\\
foutse.khomh@polymtl.ca}

\and
\IEEEauthorblockN{Teeradaj Racharak}
\IEEEauthorblockA{\textit{Tohoku University} \\
Miyagi, Japan \\
racharak.teeradaj.c3@tohoku.ac.jp}
\and
\IEEEauthorblockN{Nobukazu Yoshioka}
\IEEEauthorblockA{\textit{Waseda University} \\
Tokyo, Japan \\
nobukazu@aoni.waseda.jp}
\and
\IEEEauthorblockN{Naoyasu Ubayashi}
\IEEEauthorblockA{\textit{Waseda University} \\
Tokyo, Japan \\
ubayashi@aoni.waseda.jp}
\and
\IEEEauthorblockN{Hironori Washizaki}
\IEEEauthorblockA{\textit{Waseda University} \\
Tokyo, Japan \\
washizaki@waseda.jp}
}

\maketitle

\begin{abstract} 
Automating requirements quality analysis remains challenging because multiple, often conflicting quality attributes must be balanced while preserving stakeholder intent. Existing Large-Language-Model (LLM) approaches predominantly rely on task-oriented decomposition or implicit aggregation, limiting their ability to systematically surface and resolve cross-quality conflicts. We present \textsc{QUARE} (QUality-Aware REquirements Analysis), a multi-agent framework that takes a project description as input and formulates requirements quality analysis as structured negotiation among five quality-specialized agents (Safety, Efficiency, Green, Trustworthiness, and Responsibility), coordinated by a dedicated orchestrator. \textsc{QUARE} introduces a dialectical negotiation protocol that explicitly exposes inter-quality conflicts and resolves them through iterative proposal, critique, and synthesis. Negotiated outcomes are transformed into structurally sound KAOS goal models via topology validation and verified against industry standards through retrieval-augmented generation (RAG). We evaluate \textsc{QUARE} on five benchmark systems drawn from established RE benchmarks (MARE, iReDev) and an industrial autonomous-driving specification, spanning safety-critical, financial, and information-system domains. Results show that \textsc{QUARE} achieves 98.2\% compliance coverage (+105\% over both baselines), 94.9\% semantic preservation (+2.3 percentage points over the best baseline), and high verifiability (4.96/5.0), while generating 25--43\% more requirements than existing multi-agent RE frameworks. These findings suggest that, when using capable instruction-tuned models, architectural choices such as quality-dimension decomposition, explicit negotiation, and automated verification may contribute more to output quality than model scale alone.

\end{abstract}

\begin{IEEEkeywords}
Requirements Engineering, Multi-Agent Systems, Large Language Models, KAOS Modeling, Goal-Oriented Requirements Engineering, Requirements Negotiation
\end{IEEEkeywords}

\section{Introduction}


Modern software systems must simultaneously satisfy competing non-functional requirements such as safety, efficiency, environmental sustainability, and trustworthiness~\cite{intro_3}. As system complexity scales, particularly in safety-critical domains such as autonomous vehicles and medical devices, manually balancing these diverse quality dimensions while maintaining consistency becomes error-prone and labor-intensive~\cite{intro_4}. Empirical studies indicate that over 70\% of failed projects trace back to requirements-related issues including incompleteness, ambiguity, and unresolved stakeholder conflicts~\cite{standish2020chaos, RE}, underscoring the need for automated support in requirements quality analysis.

The requirements engineering (RE) community has long recognized the need for multi-perspective analysis. Goal-oriented methods such as KAOS (Knowledge Acquisition in Automated Specification)~\cite{vanlamsweerde2009requirements} provide obstacle analysis for reasoning about conflicts, the Delphi method~\cite{grunbacher2006multi} supports iterative stakeholder convergence, and the viewpoints framework~\cite{NewRef_Finkelstein1992} models potentially conflicting perspectives. However, these mechanisms have not been operationalized within LLM-based pipelines. Recent LLM-based multi-agent RE frameworks such as MARE~\cite{jin2024mare} and iReDev~\cite{jin2025iredev} have demonstrated the feasibility of end-to-end automation, but they decompose agents by engineering task or knowledge role rather than by quality dimension, and resolve conflicts through implicit convergence or external human intervention. As a result, cross-quality trade-offs are absorbed rather than explicitly surfaced, categorized, and resolved. For example, in an autonomous driving system, a safety agent may require sensor fusion latency $\leq 500\,\mathrm{ms}$ for thorough fault detection, while an efficiency agent demands $\leq 30\,\mathrm{ms}$ for real-time planning, resulting in a 16.7$\times$ numeric discrepancy. Such conflicts cannot be resolved through simple priority override or implicit aggregation, and instead require architectural decomposition into complementary sub-requirements.

To address these limitations, we propose QUARE, a quality-oriented multi-agent framework grounded in dialogical negotiation for requirements quality analysis. The core innovation lies in decomposing requirements analysis along quality dimensions derived from ISO/IEC 25010~\cite{intro_3} and assigning each to a specialized LLM-based agent. Drawing on the iterative refinement principles of established negotiation methods such as Delphi, these agents engage in explicit, multi-round negotiation through a five-phase pipeline that identifies, categorizes, and resolves cross-quality conflicts. Given a project description as input, QUARE produces quality-analyzed requirements organized as structurally validated KAOS goal models, verified against industry standards through retrieval-augmented generation (RAG).

The main contributions of this work are as follows:
\begin{itemize}

    \item {A \textbf{negotiation-based multi-agent framework} that formulates requirements quality analysis as collaborative negotiation among quality-specialized agents with conflict resolution across quality dimensions.}

    \item An \textbf{automated pipeline for KAOS goal modeling} that transforms unstructured requirements into structurally valid, traceable goal models.

    \item A \textbf{geometry-based evaluation methodology} for multi-quality RE that addresses the limitation of scalar metrics by operationalizing requirement diversity as convex hull volume (CHV) and mean distance to centroid (MDC) in a five-dimensional quality space, complemented by per-axis balance metrics (Coverage Uniformity and Minimum Axis Coverage), phase-wise semantic drift analysis (BERTScore), and structural compliance verification.

    \item An \textbf{open-source reimplementation suite (OpenReBench)}\footnote{\url{https://anonymous.4open.science/r/OpenRE-Bench-3CF2}} containing faithful reimplementations of all compared frameworks under identical conditions, necessitated by the absence of public implementations of MARE and iReDev, along with all experimental artifacts, prompts, outputs, and evaluation scripts.
\end{itemize}

\section{Related Work}

\subsection{Traditional Requirements Quality Analysis}

Multi-perspective requirements analysis has a long history. 
NLP-based methods have automated detection of quality defects 
such as ambiguity and incompleteness~\cite{NewRef_Femmer2017}. 
Goal-oriented methods such as KAOS~\cite{vanlamsweerde2009requirements} 
provide obstacle analysis for conflict reasoning, and 
negotiation-oriented approaches including the Delphi 
method~\cite{grunbacher2006multi} and the viewpoints 
framework~\cite{NewRef_Finkelstein1992} model 
multiple stakeholder perspectives. These methods establish 
that multi-perspective conflict resolution is essential, but 
rely on manual effort and have not been operationalized within 
LLM-based pipelines.

\subsection{Argumentation and Negotiation in RE}

Dung's abstract argumentation framework~\cite{dung} 
formalized argument acceptability, and Walton and 
Krabbe~\cite{walton} characterized dialogue types 
including negotiation. Within RE, Jureta et 
al.~\cite{NewRef_Jureta2011} applied argumentation to 
justify requirements decisions, and Mirbel and 
Villata~\cite{NewRef_Mirbel2012} integrated argumentation 
into requirements modeling. LLM-based multi-agent debate 
has been explored for general reasoning~\cite{du, 
liang} and RE tasks~\cite{NewRef_REDebate2024}, but 
without RE-specific conflict categorization, convergence 
criteria, or semantic preservation guarantees. From these works, we identify three properties for effective requirements negotiation: conflict categorization, iterative refinement with convergence criteria, and resolution traceability, against which we evaluate existing frameworks in Section II-C.

\subsection{LLM-based Single- and Multi-Agent RE}

Single-agent LLM approaches~\cite{single1, single2, 
elicitron} have shown promise in requirements elicitation 
and classification, but optimize for linguistic plausibility 
rather than multi-objective trade-offs, with dominant concerns 
overshadowing less explicit ones~\cite{intro_5, levy2024same}.
Multi-agent frameworks address this partially. General-purpose 
systems such as MetaGPT~\cite{NewRef_MetaGPT} demonstrate 
role-based coordination but are not tailored to RE quality 
conflicts. MARE decomposes RE by engineering task 
with single-turn negotiation, iReDev uses knowledge-driven agents with human-in-the-loop resolution. 
Both works demonstrate feasibility, yet important gaps remain against the three criteria above. Conflict types are treated uniformly, with no distinction between resource-bound contention and logical incompatibility. Negotiation is bounded by a fixed round count with no convergence criterion, and trade-off rationale is lost after resolution. More fundamentally, because both systems partition agents by task role rather than quality dimension, cross-quality tensions are surfaced only incidentally.
QUARE addresses these gaps by grounding agent decomposition in 
ISO/IEC 25010 quality dimensions, instantiating a dialectical 
protocol with type-specific conflict resolution and 
BERTScore-based convergence, and preserving negotiation rationale 
as auditable traces.

\vspace{-1.5ex}
\section{Methodology}

\begin{figure*}[htbp]
  \centering
  \includegraphics[width=1.0\textwidth]{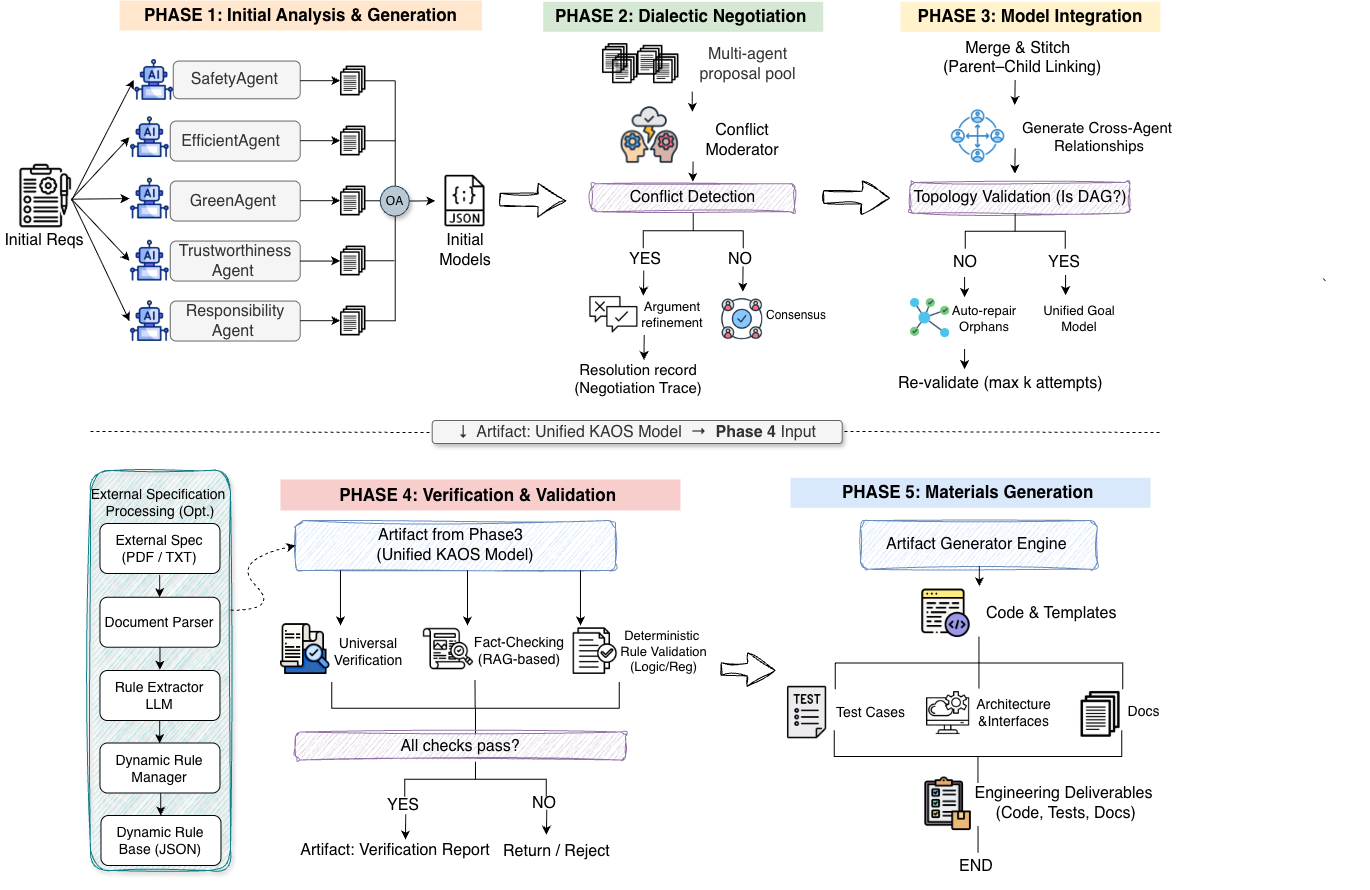}
  \caption{Overview of the QUARE framework. The pipeline comprises five sequential phases: parallel agent-based generation (Phase 1), dialectical conflict negotiation (Phase 2), KAOS goal model integration and topology validation (Phase 3), multi-layer verification and compliance checking (Phase 4), and standardized engineering materials generation (Phase 5).}
  \label{fig:overview}
\end{figure*}
The following research questions (RQs) guided the research in our work:

RQ1: Does specialized multi-agent collaboration improve the comprehensiveness, diversity, and dimensional balance of generated requirements compared to single-agent generation?

We compare the final outputs of all frameworks 
(single-agent, multi-agent without negotiation, MARE, iReDev, 
and QUARE). The multi-agent without negotiation ablation 
(Phase~1 only) isolates the contribution of agent specialization 
from negotiation. Comprehensiveness is operationalized as the 
total number of unique requirements and geometric coverage in a 
five-dimensional quality space (CHV, MDC). Dimensional balance 
is measured by Coverage Uniformity (CU) and Minimum Axis 
Coverage (MAC).

RQ2: How does multi-agent negotiation affect conflict resolution effectiveness while preserving the original semantic intent of requirements?

We trace requirements across Phases 1$\rightarrow$2$\rightarrow$3 to assess negotiation effectiveness. Conflict resolution is measured by the Conflict Resolution Rate (CRR). Semantic preservation is quantified using 
set-level BERTScore F1 between phase outputs (Algorithm~1). 
All three multi-agent frameworks are compared under identical 
conditions.

RQ3: How effectively does the framework generate structurally valid and industry-compliant requirement models?

We evaluate Phase 3 and Phase 4 outputs. Structural validity is assessed through DAG topology verification and logical consistency ($S_{\mathrm{logic}}$). Industry compliance is measured by RAG-based clause entailment coverage against applicable standards (ISO 26262, ISO 27001) and ISO/IEC/IEEE 29148 quality scores.

\subsection{Multi-Agent Framework}
\label{sec:architecture}

To address the multi-dimensional nature of requirements quality, we adopt a role-based multi-agent framework comprising six distinct roles: five quality-specialized agents and one orchestration role (Table~\ref{tab:agent_roles}). The quality-specialized agents represent core non-functional concerns: Safety, Efficiency, Sustainability, Trustworthiness, and Responsibility, while a dedicated Orchestrator coordinates these agents and integrates their outputs.

\begin{table}[tbp]
\caption{Specialized Agents within QUARE}
\label{tab:agent_roles}
\centering
\begin{tabularx}{\columnwidth}{l l X}
\hline
\textbf{Agent} & \textbf{Quality Concern} & \textbf{Core Responsibility} \\ \hline
Safety Agent &
Safety \& Reliability &
Hazard identification and mitigation \\

Efficiency Agent &
Performance &
Resource optimization and latency \\

Green Agent &
Sustainability &
Energy efficiency and carbon footprint \\

Trustworthiness Agent &
Security \& Privacy &
Data protection and access control \\

Responsibility Agent &
Ethical \& Compliance &
Regulatory and social responsibility \\

Orchestrator Agent &
System Integration &
Multi-agent coordination and workflow orchestration \\ \hline
\end{tabularx}
\end{table}

The five specialized agents are instantiated using a shared LLM backbone but are functionally isolated through dedicated system prompts. Each prompt comprises three components: (1)~a role definition specifying the quality dimension and prioritization criteria; (2)~a chain-of-thought task instruction directing the agent to analyze input, identify quality-relevant concerns, and formulate requirements; and (3)~an output schema enforcing JSON-structured requirements with fields for goal ID, description, quality dimension, KAOS level, and rationale. All agents use zero-shot prompting.

The selection of these five quality dimensions is grounded in the ISO/IEC 25010 SQuaRE (Systems and Software Quality Requirements and Evaluation) quality model~\cite{square}, which defines a hierarchical taxonomy of eight product quality characteristics. Safety and Reliability correspond to the Reliability and Safety characteristics of ISO 25010; Efficiency maps to Performance Efficiency; Trustworthiness subsumes the Security characteristic; and Responsibility captures Maintainability and compliance-oriented aspects formalized in recent extensions for AI systems~\cite{responsibility}. The Sustainability (Green) dimension, while not explicitly present in the original standard, reflects the growing consensus on green software~\cite{green} and has been proposed as an extension in recent literature~\cite{intro_5}. This grounding ensures principled, non-overlapping agent decomposition with established measurement criteria. We note that MARE and iReDev adopt different decomposition strategies (by engineering task and knowledge role, respectively), which serve complementary purposes but do not target quality-dimension coverage.

The rationale for this specialization is twofold. First, it mitigates prompt overloading, where an LLM may overlook attribute-specific constraints when exposed to an overly broad prompt \cite{levy2024same, intro_5}. Second, it deliberately induces structured disagreement among agents with competing objectives (e.g., Safety vs.\ Efficiency).

\subsection{The Five-Phase Requirement Pipeline}
\label{sec:pipeline}

To illustrate the pipeline concretely, we trace a running example drawn from the AD benchmark system. 

The operational workflow is organized into a five-phase pipeline (Figure~\ref{fig:overview}), ensuring systematic transition from raw project descriptions to formalized goal models. The pipeline separates concerns across phases: diversity-oriented generation (Phase~1), semantic reconciliation via negotiation (Phase~2), structural consolidation through model integration (Phase~3), constraint enforcement through verification (Phase~4), and final materialization (Phase~5). Negotiation and verification serve distinct roles: conflicts are resolved at the semantic level before structural integration. At the same time, compliance checking is enforced as an independent validation gate that does not alter negotiated intent.

\textbf{Phase 1: Parallel Generation} 

Five specialized agents operate in parallel, each processing the same project description using quality-oriented system prompts to produce diverse candidate requirements.

\smallskip\noindent\textit{Running Example: } Given the AD project description as input, the Safety Agent generates a sensor fusion requirement: Sensor fusion latency shall not exceed 500\,ms to enable dual-channel verification per ISO 26262-5 ASIL D.'' Independently, the Efficiency Agent generates: Sensor fusion shall complete within 30\,ms to support 10\,Hz real-time planning cycles.'' The two requirements address the same subsystem but impose a 16.7× numeric gap, which is surfaced in Phase~2.

\textbf{Phase 2: Dialectical Negotiation} 

The coordinator identifies semantic overlaps and logical
contradictions. Agents then engage in multi-round deliberation
following a thesis, antithesis, and synthesis structure to refine
or merge conflicting goals. Each round concludes with pairwise
cross-agent validation. The negotiation protocol is detailed in Section~\ref{sec:negotiation}.

\smallskip\noindent\textit{Running Example: } The Conflict Coordinator computes a cosine similarity of $0.91$ between the two sensor fusion requirements, exceeding $\tau = 0.85$, and the LLM classifier labels the pair as a resource-bound conflict. Over three negotiation rounds, the initial $16.7\times$ latency gap is progressively narrowed: the Safety Agent's original $500\,\mathrm{ms}$ constraint is decomposed into a synchronous fast path ($\leq 30\,\mathrm{ms}$, $\geq 99\%$ of cases) and an asynchronous thorough path ($\leq 500\,\mathrm{ms}$, anomaly-triggered), supplemented by a fallback mechanism ensuring planning continuity. The full round-by-round negotiation trace is presented as a benchmark system case in Section~V-B2.

\textbf{Phase 3: Integration \& Topology Validation} 

KAOS is a goal-oriented RE methodology that organizes requirements into a three-level goal hierarchy: strategic goals capture high-level stakeholder objectives, tactical goals specify system-level functions, and operational goals define implementable actions. Goals are linked by AND/OR refinement relations \cite{vanlamsweerde2009requirements}. QUARE adopts this structure as its integration target, transforming the negotiated requirements from Phase 2 into a unified KAOS goal model.

Fragmented requirements are synthesized into a coherent KAOS goal model through three steps. First, semantic deduplication merges requirements whose embedding similarity exceeds a threshold, eliminating near-identical goals generated independently by different agents. Second, cross-agent parent–child stitching establishes refinement links between goals from different agents (e.g., linking a safety sub-goal to a system-level strategic goal proposed by another agent), producing a unified goal hierarchy. Third, the resulting graph is validated as a directed acyclic graph (DAG). Cycles are prohibited because they would imply that a goal depends on its own sub-goals, undermining traceability. Structural corrections preserve negotiated intent from Phase~2.

{

\smallskip

\noindent\textit{Running Example:} The three negotiated sub-requirements from Phase 2 are organized into a KAOS hierarchy: a strategic goal (``Maintain vehicle safety under all operating conditions"), a tactical goal synthesized to reconcile safety and efficiency concerns (``Ensure safe and efficient sensor fusion"), and the three operational goals TG2.1–TG2.3 as AND-refinements. Cross-agent stitching places TG2.1 (Efficiency Agent) and TG2.2 (Safety Agent) under the same tactical parent, DAG validation confirms no circular dependencies.

}

\textbf{Phase 4: Verification}

Multi-layer validation consists of three complementary checks. First, deterministic rule checking enforces syntactic and structural constraints. For example, each operational goal must be linked to at least one tactical parent goal. Second, RAG-based compliance verification retrieves relevant clauses from domain standards (e.g., ISO 26262 and ISO 27001) and evaluates whether generated requirements satisfy the corresponding regulations, while identifying unsupported claims as potential hallucinations. Third, compliance coverage measures the extent to which the generated requirement set addresses the applicable standard clauses. This phase enriches the requirements with structural annotations while preserving their original intent.

\smallskip\noindent\textit{Running Example: } Deterministic rule checking verifies that all three operational goals trace to a tactical parent. RAG-based compliance verification retrieves ISO~26262-5 Clause~7 on fault detection timing and confirms that TG2.1 and TG2.2 jointly satisfy the clause. The fast-path latency requirement ($\leq 30\,\mathrm{ms}$) and the thorough-path coverage requirement ($\leq 500\,\mathrm{ms}$ for anomaly cases) together address both the real-time and safety-critical constraints specified by the standard.

\textbf{Phase 5: Standardized Output Generation}

Phase 5 further supports the generation of formatted deliverables, such as JSON representations, GSN XML specifications, and documentation artifacts. Because this functionality is not evaluated in the present study, we exclude a detailed description of the phase.

\subsection{Dialectical Negotiation Protocol}
\label{sec:negotiation}

The protocol instantiates a structured dialectical negotiation inspired by formal argumentation~\cite{dung}, adapted to multi-quality requirements negotiation. We adopt \emph{dialectical} as the primary term to emphasize the thesis--antithesis--synthesis structure. The protocol is also \emph{dialogical} in the sense of Walton and Krabbe~\cite{walton}, as it unfolds through multi-party dialogue. Each round consists of a thesis (agent proposal), antithesis (peer critique citing constraint violations), and synthesis (moderated resolution), with explicit engineering termination and traceability conditions. Unlike implicit aggregation or heuristic-based debate, the protocol treats conflict resolution as a primary operation, prioritizing semantic preservation alongside transparent trade-off management.

\paragraph{Conflict Detection and Categorization}
We define a conflict as a state in which the satisfaction of an agent-specific constraint $C_i$ undermines the attainment of another constraint $C_j$. Conflicts fall into two classes. The first concerns \emph{resource-bound conflicts}, which arise when quality attributes compete for finite resources (e.g., safety redundancy vs.\ power minimization). The second consists of \emph{logical incompatibilities}, in which requirements impose mutually exclusive system states (e.g., encryption latency vs.\ response-time constraints).

A specialized Conflict Coordinator mediates conflict detection in two stages. First, the coordinator computes pairwise cosine similarities over \texttt{bert-base-uncased} embeddings of agent-generated requirements and flags pairs whose similarity exceeds $\tau = 0.85$ as potential overlaps. We select \texttt{bert-base-uncased} for its low inference latency, as this stage serves as coarse retrieval prior to LLM-level classification. Second, the LLM classifies each flagged pair as (a)~redundant, (b)~resource-bound conflict, or (c)~logical incompatibility. Separating embedding-level retrieval from LLM-level classification allows the coordinator to surface multi-dimensional contradictions that heuristic merging would absorb.

\paragraph{Dialectical Loop}
The protocol operates through three coordinated stages. In the first stage, agents independently analyze requirement pairs to detect semantic overlaps and logical contradictions, establishing a conflict registry that prioritizes high-severity incompatibilities. The second stage initiates dialectical rounds, in which agents engage in iterative proposal, critique, and synthesis cycles, capped at 3 rounds to ensure termination. The third stage validates the resolution outcome, verifying that all registered conflicts have been addressed and that no new contradictions have been introduced.

In each iteration, a focus agent $A_i$ proposes a candidate requirement set $R_i$ reflecting its primary quality concern. Peer agents evaluate $R_i$ against their own constraints and articulate rationalized critiques. These critiques are integrated through neutral moderation, resolving conflicts either by jointly satisfying multiple constraints or by applying project-specific priority weights. Unresolved conflicts after three rounds are escalated to priority-weighted integration in Phase~3, preserving all conflict rationale in the negotiation trace.

By design, the negotiation phase records conflict signals, resolution states, and trade-off rationale as structured metadata rather than directly rewriting requirement text. Textual modifications are deferred to Phase~3, where deduplication and structural harmonization are applied.

We formalize this process as follows. Let $\mathcal{A} = \{A_1, \ldots, A_n\}$ denote specialized agents with quality objectives $Q_i$ and priority weights $w_i$. The negotiation approximates:
\begin{equation}
r^* = \arg\max_r \sum_{i=1}^{n} w_i \cdot Q_i(r)
\end{equation}
The weights $w_i$ are initialized uniformly ($w_i = \frac{1}{n}$) and can be overridden by project-specific configuration. At round $k$, requirement $r_k$ is updated via:
\begin{equation}
r_{k+1} = G(r_k, \text{critique}(A_j, r_k))
\end{equation}
where $G$ denotes the synthesis function integrating peer critiques into a revised proposal, the loop terminates when semantic similarity between successive rounds exceeds $1 - \epsilon$ (measured by BERTScore) or $N = 3$ rounds is reached, ensuring both convergence and bounded computation.

Eq.~(2) is order-sensitive in principle. QUARE mitigates this via round-robin scheduling (each agent as focus agent once per round) with coordinator-aggregated critiques. Experiments across three random seeds show low variance, indicating that multi-round synthesis substantially reduces order sensitivity.

\section{Experimental Setup}

\subsection{Benchmark Systems}
We evaluate QUARE using five benchmark systems spanning safety-critical, financial, and information system domains (Table~\ref{tab:casestudies}). The Library and ATM systems are drawn from the MARE benchmark, while the RollCall and Bookkeeping systems are adopted from iReDev. We additionally include an Autonomous Driving (AD) case derived from industrial specifications~\cite{baidu}, representing a safety-critical system with rich cross-quality constraints. Across all five benchmark systems, specialized agents are activated to ensure comprehensive quality coverage and enable fair comparisons across systems with varying quality emphases.

All inputs are project-level descriptions rather than pre-existing requirement sets. The ``Reqs'' column in Table~\ref{tab:casestudies} reports the number of seed requirements contained in each input document. The requirement counts reported in the evaluation (e.g., Table~III) refer to requirements that agents derive from these inputs through quality-oriented analysis, not the input seeds themselves.


\begin{table}[tbp]
\caption{Summary of Benchmark Systems}
\label{tab:casestudies}
\centering
\footnotesize
\setlength{\tabcolsep}{2.8pt}
\renewcommand{\arraystretch}{0.95}

\begin{tabular}{llllcc}
\toprule
\textbf{System} & \textbf{Domain} & \textbf{Focus} &
\textbf{Input} & \textbf{Words} & \textbf{Reqs.} \\
\midrule
AD          & Automotive & Safety      & Structured & 323 & 10 \\
ATM         & Finance    & Security    & Narrative  & 165 & 10 \\
Library     & Management & Scalability & Compact    & 72  & 5  \\
RollCall    & Management & Logic       & Compact    & 58  & 5  \\
Bookkeeping & Finance    & Integrity   & Compact    & 64  & 5  \\
\bottomrule
\end{tabular}
\end{table}

\subsection{Agent and Model Configuration}

The similarity threshold $\tau = 0.85$ was selected based on a 
pilot sweep over $\tau \in \{0.75, 0.80, 0.85, 0.90, 0.95\}$ 
on the AD benchmark system, values below 0.80 produced excessive 
false positives, while values above 0.90 missed genuine conflicts. All experiments use \texttt{gpt-4o-mini-2024-07-18} at 
temperature 0.7, balancing linguistic diversity with structural 
consistency. This choice serves two purposes: sharing a single 
backbone ensures that inter-agent disagreements arise solely 
from prompt-induced role specialization, isolating the 
architectural contribution, and the model's low per-token cost makes 180 
experiment runs economically feasible while providing sufficient 
reasoning capacity, as evidenced by the compliance and semantic 
preservation scores. We acknowledge that heterogeneous or larger 
models could amplify inter-agent diversity and plan to investigate 
this in future work.

Agents employ ChromaDB for RAG, grounding their 
reasoning in external standards including ISO~26262 and 
ISO~27001. Generated goal models conform to a three-level KAOS 
hierarchy (Strategic, Tactical, and Operational). For conflict 
detection, pairwise cosine similarities are computed using 
\texttt{bert-base-uncased} due to its low latency. 
Evaluation-stage semantic preservation uses BERTScore with 
contextual embeddings. The negotiation process is capped at 
three rounds per agent, determined empirically: over 90\% of 
resolvable conflicts converged by Round~2, with Round~3 yielding 
marginal refinement. Prompts were developed through iterative 
refinement on the AD system and applied without modification to 
the remaining four systems, full prompts are provided in the 
replication package. Each experiment is executed three times with 
different random seeds, and results are averaged to mitigate 
non-determinism.

\subsection{Baselines and Comparison Methods}
We compare against four baselines spanning increasing
levels of multi-agent coordination.

\textbf{Single-agent baseline.} A single LLM instance
performs RE in a single pass, without role specialization
or iterative refinement.

\textbf{Multi-agent without negotiation.} Five specialized
agents operate in parallel (corresponding to Phase~1 of
our pipeline), with outputs directly aggregated without
conflict resolution. This ablation isolates the effect of
dialectical negotiation (RQ2).

\textbf{MARE.} A task-specialized multi-agent framework
with five agents organized around a 9-action pipeline.
Agents collaborate through a shared workspace with
single-turn negotiation, decomposing RE by \textit{task
type} rather than by quality dimension.

\textbf{iReDev.} A knowledge-driven multi-agent framework
with six agents and a 17-action workflow. In our
reimplementation, human-in-the-loop interactions are
operationalized via LLM-based surrogate agents with
deterministic fallback, enabling full automation and
reproducibility while acknowledging that this may not
fully capture the quality of real human feedback.

The three frameworks differ in decomposition strategy, specifically by engineering task (MARE), knowledge role (iReDev), and quality dimension (QUARE), enabling evaluation of how decomposition philosophy affects output properties (discussed in Section VI-A). Since neither MARE nor iReDev provides public implementations, we reimplemented both within the shared OpenReBench suite following their published descriptions, adopting conservative interpretations where details were underspecified. All frameworks share the model and parameter configuration described in Section IV-B; experiments span four settings and three random seeds (101, 202, 303), yielding 180 total runs.

\subsection{Evaluation Metrics}
We define evaluation metrics aligned with our three RQs, covering requirement diversity, negotiation effectiveness,
and structural validity under regulatory constraints.

\paragraph{RQ1: Coverage and Diversity}
This RQ examines whether the generated requirements
span diverse quality concerns,
rather than clustering around a small set of dominant attributes.
To this end, we embed requirements into a five-dimensional quality space
$\mathcal{Q} = \mathbb{R}^{5}$,
where each axis corresponds to one of the five quality attributes
managed by the specialized agents:
Safety ($q_1$), Efficiency ($q_2$),
Sustainability ($q_3$), Trustworthiness ($q_4$),
and Responsibility ($q_5$).

Each requirement $r_i$ is projected into $\mathcal{Q}$ by scoring its
relevance to each quality axis on a $[0,1]$ scale, using a zero-shot
LLM classifier (\texttt{gpt-4o-mini}) prompted with axis-specific rubrics
(e.g., the Safety rubric: \textit{`Rate how directly this requirement
addresses hazard identification, fault tolerance, or safety-critical
behavior'}). All five axes are evaluated in a single call, producing
a five-dimensional relevance vector. The classifier is identical to the one used in the blind re-labeling analysis in
Section VII-A. Full rubrics are provided in the
replication package. Let $P = \{r_1, \dots, r_m\} \subset \mathcal{Q}$
denote the resulting set of projected vectors, treated as a point set in the geometric sense for the coverage measures below.
Traditional metrics such as precision and recall require a canonical reference set, which does not exist for open-ended requirements generation. Scalar counts can also mask dimensional imbalance, as a system producing many safety requirements but none for sustainability may still score well in aggregate. Geometry-based measures require no reference set and jointly characterize coverage extent (CHV) and dispersion (MDC) across all five dimensions. Scalar metrics such as per-axis averages can mask dimensional imbalances (e.g., high safety but neglected sustainability may still yield a high aggregate score). Geometry-based measures treat requirements as a point cloud in quality space, capturing coverage extent (CHV) and spread (MDC) without collapsing dimensions.

We quantify coverage and diversity using two geometry-based measures.
The first is the convex hull volume (CHV),
defined as the volume of $\mathrm{conv}(P)$ in $\mathbb{R}^{5}$,
computed via standard Delaunay triangulation. Higher CHV values indicate broader coverage across quality dimensions.
The second metric is the mean distance to centroid (MDC),
which captures the dispersion of requirements within the quality space.
Formally, let $\bar{q} = \frac{1}{m}\sum_{i=1}^{m} r_i$ denote the centroid of $P$,
and define
\begin{equation}
    \mathrm{MDC} = \frac{1}{m}
    \sum_{i=1}^{m} \| r_i - \bar{q} \|_2 .
\end{equation}
A higher MDC reflects greater dispersion
and reduced redundancy among generated requirements.

To complement the geometric measures above, which capture overall spread in quality space, we introduce two per-axis distribution metrics that characterize dimensional balance. For multi-agent frameworks whose architecture decomposes requirements by quality dimension, let $n_k$ ($k=1,\dots,5$) denote the number of requirements assigned to axis $k$, and let $\bar{n}=\frac{1}{5}\sum_{k} n_k$.
Coverage Uniformity (CU) is defined as the standard deviation of per-axis counts:
\begin{equation}
    \mathrm{CU} = \sqrt{\frac{1}{5}\sum_{k=1}^{5}(n_k - \bar{n})^2} .
\end{equation}
Lower CU indicates more balanced coverage across quality dimensions.
Minimum Axis Coverage (MAC) is defined as $\mathrm{MAC} = \min_k\, n_k$, capturing the worst-case dimensional guarantee. Higher MAC ensures that no single quality dimension is systematically neglected.
CU and MAC require per-axis assignment of requirements to quality dimensions. Multi-agent frameworks provide this naturally through agent specialization, whereas the single-agent baseline lacks such decomposition, we therefore omit CU and MAC for the single-agent baseline.

\paragraph{RQ2: Negotiation Effectiveness}
This RQ evaluates whether dialectical negotiation
successfully resolves conflicts
while preserving original stakeholder intent.
We measure negotiation effectiveness along three complementary dimensions.
First, the CRR measures the proportion of detected contradictions that negotiation successfully resolves. Second, we assess semantic preservation using BERTScore.
Finally, logical consistency ($S_{\text{logic}}$) is evaluated
through ISO~29148-based checks,
detecting numeric or boolean inconsistencies
across the resulting requirement set.

To quantify semantic preservation at the set level, we extend 
BERTScore from individual sentence pairs to requirement sets 
via optimal bipartite matching. Let $R^{(a)}=\{r^{(a)}_1, 
\ldots, r^{(a)}_m\}$ and $R^{(b)}=\{r^{(b)}_1, \ldots, 
r^{(b)}_n\}$ denote the requirement
sets produced at phases $a$ and $b$ (e.g., Phase~1 vs.\ Phase~3).
We compute a pairwise similarity matrix
$S_{ij} = \mathrm{BERTScoreF1}(r^{(a)}_i, r^{(b)}_j)$
using \texttt{bert-base-uncased}, then solve a maximum-weight bipartite matching
to obtain a one-to-one alignment that accounts for set-size differences via
dummy padding. Algorithm~\ref{alg:set_level_preservation} summarizes the
procedure, the resulting score $P \in [0,1]$ decreases when requirements are
rewritten (lower pairwise similarity) or when the set size changes (unmatched
dummy nodes contribute zero similarity).

\begin{algorithm}[t]
\caption{Set-level semantic preservation via optimal matching}
\label{alg:set_level_preservation}
\begin{algorithmic}[1]
\Require Requirement sets $\mathcal{R}^{(a)}=\{r^{(a)}_i\}_{i=1}^{m}$,
         $\mathcal{R}^{(b)}=\{r^{(b)}_j\}_{j=1}^{n}$
\Ensure  Preservation score $P \in [0,1]$
\State Compute pairwise similarity
       $S_{ij} \gets \mathrm{BERTScoreF1}(r^{(a)}_i, r^{(b)}_j)$
\State $N \gets \max(m,n)$
\State Pad $S$ to $S' \in \mathbb{R}^{N \times N}$ with dummy rows/cols;
       set dummy similarities to $0$
\State Solve $\pi^* \gets \mathrm{HungarianMax}(S')$
       \Comment{max-sum assignment}
\State $P \gets \frac{1}{N}\sum_{i=1}^{N} S'_{i,\pi^*(i)}$
\State \Return $P$
\end{algorithmic}
\end{algorithm}

\paragraph{RQ3: Structural Validity and Compliance}
This RQ assesses whether negotiated requirements
can be reliably mapped to formal goal models
and validated against regulatory standards.
Structural validity is evaluated by verifying
acyclicity and parent--child integrity
in the resulting KAOS goal graph.
Compliance coverage measures
the proportion of applicable regulatory clauses
(e.g., ISO~26262) satisfied by the generated model.
In addition, terminology consistency ($S_{\text{term}}$)
is assessed on a 1--5 scale following ISO~29148,
capturing the semantic coherence of concepts across the model
and ensuring unified naming
(e.g., harmonizing ``client'' and ``user'').

To evaluate industry alignment, we measure \emph{compliance coverage} against
domain standards (e.g., ISO~26262 for AD; ISO~27001 for security-centric systems).
Compliance coverage is evaluated through a RAG-based clause 
entailment pipeline. Let $C$ denote a corpus of standard 
clauses. For each applicable clause $c \in \mathcal{C}_{\text{app}}$, we retrieve the
top-$k$ relevant requirements $\{r_1,\dots,r_k\} \subseteq \mathcal{R}$ via
embedding similarity, along with supporting standard context (neighboring
paragraphs, definitions) to form an evidence bundle
$E(c) = \{\text{clause text}, \text{context}, r_1,\dots,r_k\}$.
An independent verifier LLM then predicts an entailment label
$y(c) \in \{\textsc{Satisfied}, \textsc{Partially}, \textsc{NotSatisfied}\}$,
together with the best supporting requirement, a short rationale, and an
extractive citation. A clause is counted as \emph{covered} when
$y(c) \in \{\textsc{Satisfied}, \textsc{Partially}\}$.
Compliance coverage is defined as the proportion of applicable clauses that are
covered:
\begin{equation}
\mathrm{Coverage}(\mathcal{R}) =
\frac{1}{|\mathcal{C}_{\text{app}}|}
\sum_{c \in \mathcal{C}_{\text{app}}} \mathbf{1}\!\big[y(c) \in
\{\textsc{Satisfied},\textsc{Partially}\}\big].
\label{eq:compliance_coverage}
\end{equation}
We report coverage averaged over three random seeds. To reduce verifier variance,
we fix the verifier model and decoding parameters and apply a self-consistency
majority vote over $n{=}3$ repeated judgments per clause.

\section{Results and Analysis}
\label{sec:results}
In this section, we present results validating the effectiveness of QUARE. Our analysis follows three RQs, demonstrating how the framework progresses from diverse generation to consistent, engineering-ready models.

\subsection{Results for RQ1: Coverage and Diversity}

RQ1 examines whether specialized multi-agent collaboration improves the coverage, diversity, and dimensional balance of generated requirements. We compare the final outputs of all five configurations (single-agent, multi-agent without negotiation, MARE, iReDev, and QUARE), where the multi-agent-without-negotiation ablation isolates the contribution of agent specialization from subsequent pipeline stages. All geometric metrics (CHV and MDC) are computed in the original five-dimensional quality space $Q=\mathbb{R}^{5}$, with PCA used solely for visualization.

\begin{table}[t]
\centering
\caption{Requirement Coverage and Diversity (averaged across 5 benchmark systems)}
\label{tab:req_coverage_diversity}
\resizebox{\columnwidth}{!}{%
\begin{tabular}{lccccc}
\toprule
Framework & Req.\ Count ($\uparrow$) & CHV ($\uparrow$) & MDC ($\uparrow$) & CU ($\sigma$, $\downarrow$) & MAC ($\uparrow$) \\
\midrule
Single-agent & 14.2 & 2.8 & 0.715 & ---   & ---  \\
MARE         & 24.4 & 4.8 & 0.835 & 0.30  & 4.6  \\
iReDev       & 28.1 & 6.4 & 0.705 & 0.45  & 5.1  \\
QUARE        & 35.0 & 4.3 & 0.673 & \textbf{0.20} & \textbf{6.7} \\
\bottomrule
\end{tabular}%
}
\vspace{0.3em}

\raggedright\footnotesize
\textbf{Req.\ Count}: total number of unique generated requirements.
\textbf{CHV}: convex hull volume in 5-D quality space (values $\times 10^{-3}$).
\textbf{MDC}: mean distance to centroid.
\textbf{CU}: coverage uniformity (Eq.~4); lower is more balanced. 
\textbf{MAC}: minimum axis coverage, i.e., the lowest per-axis requirement count; higher values indicate stronger worst-case coverage.
CU and MAC are not applicable to the single-agent baseline, which lacks quality-dimension decomposition.
\end{table}



The quantitative results in Table~\ref{tab:req_coverage_diversity} confirm that multi-agent outputs achieve broader coverage than single-agent baselines across all five evaluation subjects. We provide detailed per-case breakdowns and visualizations in the supplementary material. The multi-agent withoutnnegotiation ablation (Phase 1 only) yields similar coverage geometry to QUARE's final output, as negotiation preserves rather than transforms the requirement set (Phase 2 vs. Phase 1 BERTScore: 100.0\%), its detailed contribution is isolated in Section V-E.

The three multi-agent systems exhibit complementary strengths across coverage metrics, reflecting distinct architectural philosophies. QUARE generates the most requirements (35.0 avg, +43\% over MARE, +25\% over iReDev), driven by five quality-specialized agents that systematically surface dimension-specific concerns. iReDev achieves the highest CHV (0.0064, +48\% over QUARE), indicating the broadest quality-space exploration through its 6-agent knowledge-driven pipeline. MARE leads MDC (0.835, +24\% over QUARE), producing the most dispersed, non-redundant distribution.

While QUARE's CHV and MDC are lower than those of iReDev 
and MARE respectively, this reflects an architectural trade-off 
inherent to quality-specialized decomposition. Because each agent 
focuses on a single quality dimension, generated requirements 
naturally cluster along their assigned axes, reducing geometric 
spread but improving dimensional balance and worst-case coverage. 
QUARE achieves the lowest CU (0.20 vs.\ 0.30 for MARE and 
0.45 for iReDev) and the highest MAC (6.7 vs.\ 4.6 and 5.1), 
meaning that even its least-represented dimension receives more 
requirements than the worst-case axis of either baseline. This 
balanced coverage directly supports compliance outcomes: in the 
AD case, the Safety agent's requirements achieve 91.1\% 
ISO~26262 coverage versus MARE's 7.8\% and iReDev's 10.0\%, 
indicating that domain-specific emphasis is expressed through semantic depth rather than raw count. These distinct coverage profiles reflect the underlying decomposition philosophies, which we analyze further in Section~VI-A.

\begin{rqbox}{Answer to RQ1}
Multi-agent collaboration yields up to 2.5$\times$ more requirements, 129\% higher CHV, and 17\% higher MDC compared to single-agent generation. Among multi-agent frameworks, each architectural philosophy produces distinct coverage characteristics: QUARE generates the most requirements (35.0, +43\% over MARE) with the most balanced dimensional coverage (CU\,=\,0.20 vs.\ 0.30/0.45) and strongest worst-case axis guarantee (MAC\,=\,6.7 vs.\ 4.6/5.1); iReDev achieves the broadest quality-space exploration (CHV\,=\,0.0064); and MARE produces the most dispersed distribution (MDC\,=\,0.835). QUARE's balanced coverage translates into compliance advantages examined in RQ3 (98.2\% vs.\ $\sim$47\%).

\end{rqbox}


\subsection{Results for RQ2: Negotiation Effectiveness and Semantic Preservation}

RQ2 investigates how dialectical negotiation resolves conflicts while preserving semantic intent. We perform a phase-wise trajectory analysis across four milestones: Phase 1 (Raw Generation), Phase 2 (Negotiated Alignment), Phase 3 (Unified Integration), and Phase 4 (Verified Output), comparing all three
frameworks under identical conditions.

\subsubsection{Quantitative Results}
Table~\ref{tab:semantic_preservation} shows that QUARE achieves the highest
semantic preservation across all cases (94.9\% avg BERTScore between Phase 3 and Phase 1 outputs, +5.9 percentage points over MARE, +2.3 percentage points over iReDev), with perfect Phase 2 vs. Phase 1 preservation (100.0\%). Table \ref{tab:negotiation_summary} further shows that this result is achieved despite conducting the deepest negotiation among all three systems (16.5 steps avg).

The lower CRR of QUARE (25.0\% vs.\ MARE's 66.7\%)
reflects a difference in detection granularity rather than resolution
quality. MARE employs coarser conflict detection, marking fewer total
conflicts and resolving most through simple priority override. QUARE's
two-stage detection (embedding similarity + LLM classification) surfaces
more conflicts, including architectural inconsistencies and cross-level
semantic misalignments that coarser detection would miss. The larger
denominator lowers the resolution rate. Unresolved conflicts are not
discarded but handled through priority-weighted integration in Phase~3,
as reflected in QUARE's higher overall semantic preservation
(Table~\ref{tab:semantic_preservation}). We discuss the broader implications of this detection-vs-resolution trade-off in Section VI-B.

\begin{table}[t]
\centering
\caption{Semantic Preservation (BERTScore F1, Phase 3 vs. Phase 1, $\uparrow$)}
\label{tab:semantic_preservation}
\begin{tabular}{lccc}
\toprule
Benchmark systems & MARE & iReDev & QUARE \\
\midrule
AD          & 88.4 & 92.7 & 94.8 \\
ATM         & 89.6 & 92.0 & 95.5 \\
Library     & 87.7 & 92.6 & 94.8 \\
RollCall    & 88.8 & 93.0 & 94.5 \\
Bookkeeping & 90.5 & 92.9 & 94.8 \\
\midrule
Average     & 89.0 & 92.6 & \textbf{94.9} \\
\bottomrule
\end{tabular}
\end{table}

\begin{table}[t]
\centering
\caption{Negotiation Process Summary}
\label{tab:negotiation_summary}
\begin{tabular}{lccc}
\toprule
Metric & MARE & iReDev & QUARE \\
\midrule
Avg.\ negotiation steps       & 10.0 & 12.0 & 16.5 \\
Conflict resolution rate (\%) & 66.7 & 46.7 & 25.0$^{*}$ \\
Phase 2 vs. Phase 1 BERTScore (\%)     & 97.0  & 99.4  & 100.0 \\
\bottomrule
\end{tabular}

\vspace{0.5em}
\footnotesize
$^{*}$QUARE detects 3.2$\times$ more conflicts than MARE; unresolved conflicts are
handled via priority-weighted integration in Phase~3 (see Section~III-C).
\end{table}

\subsubsection{Illustrative Example: Multi-Agent Negotiation in Autonomous Driving}

To illustrate the dialogical protocol concretely, we examine the sensor fusion
latency conflict in the AD system, one of the highest-severity incompatibilities
detected across all five benchmark systems. During Phase~1, the SafetyAgent proposed
sensor fusion latency $\leq 500$~ms (for dual-channel verification per
ISO~26262-5 ASIL~D compliance), while the EfficiencyAgent mandated $\leq 30$~ms
(for 10~Hz real-time planning cycles per Apollo Spec~\S4.2), a 16.7$\times$ numeric gap that neither MARE's nor iReDev's
implicit aggregation mechanisms could structurally resolve.

\begin{roundbox}{Round 1: Conflict Eruption}
\textbf{SafetyAgent (TG2):} Sensor fusion latency $\leq$ 500\,ms 
for comprehensive sensor validation and fault detection per 
ISO~26262-5.\\[2pt]
\textbf{EfficiencyAgent:} \textit{Strong opposition.} 500\,ms 
consumes 5$\times$ the planning cycle budget (100\,ms target). 
Sensor fusion must complete within 30\,ms per Apollo Spec \S4.2.\\[2pt]
\textbf{Classification:} High-severity numeric incompatibility 
(resource-bound). \hfill \textcolor{red}{\textbf{UNRESOLVED}}
\end{roundbox}

\begin{roundbox}{Round 2: Architectural Refinement}
\textbf{SafetyAgent (revision):} Two-stage pipeline---fast-path 
$\leq$ 30\,ms (95\% cases), thorough-path $\leq$ 500\,ms 
(anomaly-triggered).\\[2pt]
\textbf{EfficiencyAgent:} \textit{Insufficient.} Thorough-path 
must be (1)~asynchronous and non-blocking, (2)~fast-path 
$\geq$ 99\%, (3)~fallback ensures planning continuity.
\hfill \textcolor{orange}{\textbf{PARTIAL}}
\end{roundbox}

\begin{roundbox}{Round 3: Consensus \& Convergence}
\textbf{SafetyAgent (final):} Three-component architecture:
\begin{itemize}
  \item TG2.1: Fast-path $\leq$ 30\,ms, synchronous, $\geq$99\%
  \item TG2.2: Thorough-path $\leq$ 500\,ms, async, non-blocking
  \item TG2.3: Fallback---planning uses fast-path during validation
\end{itemize}
\vspace{1pt}
\textbf{EfficiencyAgent:} Fast-path satisfies G1. Async 
thorough-path has zero planning impact. 
\hfill \textcolor{green!60!black}{\textbf{CONSENSUS}}
\end{roundbox}


The subsequent negotiation proceeded through three rounds of structured
dialogical refinement, transforming this trade-off into 
an architectural solution satisfying both agents' core constraints. This negotiation trace demonstrates how dialogical negotiation enables creative architectural solutions rather than simple compromises or
priority-based rejection. By decomposing the conflicting requirement into a
synchronous fast-path (efficiency-critical, $\leq 30$~ms, $\geq 99\%$ of cases)
and an asynchronous thorough-path (safety-critical, $\leq 500$~ms, non-blocking),
the framework satisfies both agents' constraints without sacrificing either
safety or efficiency. This architectural decomposition pattern reflects the broader negotiation behavior exhibited by QUARE across all five benchmark systems, in which conflicts are resolved by restructuring the solution space rather than conceding individual constraints.

In contrast, MARE would likely resolve this via safety priority
override, discarding the efficiency constraint, while iReDev
lacks the structured dialogical mechanism to reach the
three-component synthesis. 

\begin{rqbox}{Answer to RQ2}
QUARE achieves 94.9\% semantic preservation (+5.9 percentage 
points over MARE, +2.3 percentage points over iReDev) despite 
conducting the deepest negotiation among all three systems 
(16.5 steps avg). The dialectical protocol restructures conflicting constraints into complementary sub-requirements instead of discarding lower-priority goals, producing requirements that are both logically coherent and engineering-ready.
\end{rqbox}

\subsection{Results for RQ3: Structural Validity and Compliance}

RQ3 examines whether QUARE generates structurally valid and industry-compliant requirement models.
We evaluate outputs from Phase~3 (model integration) and Phase~4 (verification and compliance checking), comparing all three frameworks
under identical experimental conditions.

\subsubsection{Structural Validity and Internal Consistency}

As shown in Table~\ref{tab:structural_compliance}~(A), all three systems
achieve comparable structural validity: valid DAG topology and zero detected
logical contradictions ($S_{\text{logic}} = 1.000$) across all benchmark systems.
Each operational element traces to at least one tactical goal, and each
tactical goal links to a strategic objective. The key differentiator is
compliance coverage. QUARE achieves 98.2\% average compliance, a $+$105\%
improvement over both MARE (47.6\%) and iReDev (47.8\%), which score nearly
identically despite differing architectures. This gap is most pronounced in
the safety-critical AD case. The advantage stems from Phase~4
RAG-augmented verification, which maps requirements to specific standard
clauses. Neither MARE nor iReDev includes such a step, their compliance
scores reflect only incidental overlap between generated requirements and
regulatory clauses.

\begin{table}[ht]
\centering
\caption{Structural Validity and Compliance}
\label{tab:structural_compliance}

\textbf{(A) Structural Correctness}

\begin{tabular}{lccc}
\toprule
Metric & MARE & iReDev & QUARE \\
\midrule
DAG Topology Valid & $\checkmark$ & $\checkmark$ & $\checkmark$ \\
Logical Consistency ($S_{\text{logic}}$) & 1.000 & 1.000 & 1.000 \\
Compliance Coverage (\%) & 47.6 & 47.8 & \textbf{98.2} \\
\bottomrule
\end{tabular}

\vspace{1em}

\textbf{(B) ISO/IEC/IEEE 29148 Quality Scores (1--5 scale, LLM-judge)}

\begin{tabular}{lccc}
\toprule
Criterion & MARE & iReDev & QUARE \\
\midrule
Unambiguous     & 4.41 & 4.19 & 4.24 \\
Correctness     & 5.00 & 5.00 & 5.00 \\
Verifiability   & 3.95 & 3.96 & \textbf{4.96} \\
Set Consistency & 5.00 & 5.00 & 5.00 \\
Set Feasibility & 3.74 & 3.75 & \textbf{4.96} \\
\bottomrule
\end{tabular}

\vspace{0.5em}
\footnotesize
Note: Compliance coverage averages ISO~26262 (AD) and ISO~27001 (remaining cases). 
Bold indicates the best result per metric.

\vspace{1em}

\textbf{(C) Human vs. LLM-Judge Validation Results}

\begingroup
\scriptsize
\setlength{\tabcolsep}{3.2pt}
\renewcommand{\arraystretch}{0.95}
\begin{tabular}{lccc ccc}
\toprule
\multirow{2}{*}{Criterion} & \multicolumn{3}{c}{Human Eval. Mean} & \multicolumn{3}{c}{LLM Judge Mean} \\
\cmidrule(lr){2-4}\cmidrule(lr){5-7}
 & MARE & iReDev & QUARE & MARE & iReDev & QUARE \\
\midrule
Unambiguous     & 3.68 & 3.70 & 3.72 & 4.41 & 4.18 & 4.23 \\
Correctness     & 4.95 & 4.93 & 4.97 & 5.00 & 5.00 & 5.00 \\
Verifiability   & 4.00 & 4.05 & 4.27 & 3.95 & 3.95 & 4.97 \\
Set Consistency & 4.89 & 4.91 & 4.92 & 5.00 & 5.00 & 5.00 \\
Set Feasibility & 4.27 & 4.27 & 4.30 & 3.73 & 3.73 & 4.97 \\
\bottomrule
\end{tabular}
\endgroup

\end{table}

\subsubsection{ISO/IEC/IEEE 29148 Evaluation}

We evaluate industrial quality against ISO~29148 at both individual requirement
and requirement-set levels, using an LLM-based judge instantiated independently
of the generation agents to mitigate self-evaluation bias.
Table~\ref{tab:structural_compliance}~(B) presents cross-system averages.
All three systems score 5.00 on both correctness and consistency, indicating
that multi-agent pipelines reliably produce well-formed, internally coherent
requirement sets. MARE leads marginally in unambiguity (4.41), which we
attribute to its task-specialized agents' focus on specification-level phrasing.
The main advantages of QUARE appear in verifiability (4.96, +25\% over MARE)
and set feasibility (4.96, +32\% over MARE). Both gains trace to the
RAG-augmented verification phase, which enforces measurable acceptance criteria
and checks implementation feasibility against domain standards. These gains
hold across all five benchmark systems and are the quality dimensions most relevant
to industrial adoption.

To validate the LLM-as-a-judge evaluation, we conducted a blind human
evaluation on a stratified sample of 74 Phase~3 requirement instances,
covering all three frameworks and all five benchmark systems (up to ten unique
requirements per framework case pair). Two annotators independently
scored each requirement using the same five-criterion ISO/IEC/IEEE~29148
rubric, framework labels were replaced with anonymous identifiers.
As shown in Table~\ref{tab:structural_compliance}~(C), QUARE obtains the highest
human score on all five criteria. The advantage is most pronounced on
verifiability (4.27 vs.\ 4.00/4.05) and set feasibility
(4.30 vs.\ 4.27/4.27), consistent with the LLM-judge results in
Table~\ref{tab:structural_compliance}~(B). Correctness and set consistency score near 5.0 for all three
frameworks in both human and LLM evaluations, indicating that all
multi-agent pipelines produce well-formed, internally coherent
requirements. While the LLM judge assigns larger score gaps between
frameworks on verifiability and feasibility than human annotators do,
both evaluation modes rank QUARE first on all five criteria. We attribute the larger LLM-judge gap to the judge's sensitivity to explicit measurability cues (e.g., numeric thresholds, standard references) that RAG-augmented outputs produce more frequently, human annotators weigh broader contextual factors, compressing score differences. The two
human annotators showed high agreement (81.1\% exact match, 99.7\%
within-one agreement), confirming that the human scores are reliable
and can serve as an independent check on the LLM-based evaluation.

\subsubsection{Regulatory Compliance and Semantic Stability}

Phase~4 applies deterministic rule checking and standards-based validation (ISO~26262 for AD, ISO~27001 for security systems). After verification, all QUARE-generated models achieve 100\% compliance. The BERTScore between Phase~3 and Phase~4 remains unchanged across all cases, confirming that compliance checking functions as a pure validation layer rather than a transformation step, and that negotiated semantic intent is fully preserved throughout the verification pipeline.

\begin{rqbox}{Answer to RQ3}
All three systems achieve structural parity (valid DAG topology, $S_{\text{logic}} = 1.000$, perfect correctness and consistency). However, \textbf{QUARE} outperforms both MARE and iReDev in compliance coverage (\textbf{98.2\%} vs. 47.6\% and 47.8\%) and requirement quality (verifiability \textbf{4.96}, feasibility \textbf{4.96}), demonstrating that automated RAG-augmented verification is the key component for producing engineering-ready, standards-compliant artifacts.
\end{rqbox}

\subsection{Cross-Framework Synthesis}
Synthesizing the results from Tables~\ref{tab:req_coverage_diversity}--\ref{tab:structural_compliance}, QUARE leads in compliance-critical metrics (compliance coverage, verifiability, feasibility, and semantic preservation), requirement count, and dimensional balance (lowest CU, highest MAC), while MARE leads in runtime and dispersion (MDC), and iReDev achieves the broadest quality-space exploration (CHV). The three systems reflect distinct design philosophies: QUARE excels in compliance-critical and safety-critical scenarios with the most balanced quality-dimension coverage, strongest semantic preservation, and regulatory alignment at moderate cost (55.4s). MARE is optimal for rapid prototyping with the fastest runtime, but lower compliance, and iReDev occupies a knowledge-intensive middle ground with the highest CHV but at 4.8$\times$ computational cost of MARE.

\subsection{Ablation Study}

This section provides a consolidated analysis of how individual components of QUARE contribute to the overall performance observed in RQ1--RQ3. Our ablation follows a phase-wise design that incrementally evaluates the effect of each processing stage.

\textbf{Agent Specialization (Phase~1):}
CHV increases 53.6\% over single-agent baselines, confirming that diversity emerges from quality-oriented specialization. MDC decreases modestly ($-5.9\%$), consistent with the architectural trade-off discussed in Section V-A: quality-specialized agents produce requirements that cluster along their assigned axes, yielding lower geometric dispersion but stronger dimensional balance (CU = 0.20, MAC = 6.7).

\textbf{Explicit Negotiation (Phase~2):}
QUARE achieves perfect semantic 
retention during negotiation (P2 vs. P1: 100.0\%), indicating 
that the dialogical protocol resolves conflicts without overwriting 
original stakeholder intent. MARE and iReDev show slight drift 
(97.0\% and 99.4\%, respectively).

\textbf{Model Integration (Phase~3):}
Integration introduces modest 
semantic adjustment (P3 vs. P1: 94.9\% for QUARE), primarily 
from deduplication and structural harmonization rather than 
intent modification.

\textbf{Verification and Compliance Checking (Phase~4):}
Zero semantic drift (P4 vs. P3 unchanged), 
confirming that compliance checking functions as a pure validation 
layer.

The design of QUARE offers several properties relevant to 
industrial adoption. First, the quality-specialized decomposition 
is extensible to additional agents without restructuring the 
pipeline. Second, the negotiation trace provides an auditable 
record of trade-off decisions, addressing traceability requirements 
in safety-critical domains (ISO 26262, IEC 62304) and emerging 
regulatory mandates (EU AI Act, Article~11). Third, the moderate 
runtime (55.4s) and compatibility with lightweight models suggest 
feasibility for iterative development workflows

\section{Discussion}

\subsection{Architectural Trade-offs}

No single decomposition strategy dominates across all metrics. 
QUARE's quality-dimension decomposition yields the most balanced 
coverage (CU = 0.20, MAC = 6.7) and highest compliance (98.2\%), 
but lower geometric spread (CHV, MDC) than iReDev and MARE. This 
reflects a design trade-off: agents constrained to a single quality 
axis produce clustered but balanced outputs, whereas task-oriented 
(MARE) or knowledge-driven (iReDev) agents explore more broadly 
but risk neglecting individual dimensions. In regulated domains where no quality dimension may be overlooked, quality-dimension decomposition is preferable, while broader strategies may be more appropriate for early-stage exploration.

QUARE generates the most requirements (35.0 avg), but higher 
count does not automatically imply higher quality. We note that 
the additional requirements are distributed across dimensions 
(lowest CU) and traceable to standard clauses (98.2\% compliance), 
mitigating inflation concerns. This concern is further mitigated by the human evaluation reported in Table~VI(C), which confirms alignment between LLM-judge scores and human assessments across all five criteria. Crucially, balanced coverage alone does not guarantee compliance: the CU→compliance connection (0.20 → 98.2\%) depends on the downstream negotiation and verification pipeline, not on agent specialization itself.
While the Sustainability and Performance agents address 
overlapping resource concerns, they differ in scope: 
Performance targets runtime efficiency (latency, throughput), 
whereas Sustainability targets lifecycle resource impact 
(energy consumption, carbon footprint). Across all five 
benchmark systems, the two agents produced fewer than 
8\% overlapping requirements.

\subsection{Conflict Resolution}

QUARE's lower CRR (25.0\% vs.\ MARE's 66.7\%) reflects detection 
depth rather than resolution weakness: QUARE surfaces 
3.2$\times$ more conflicts, mechanically lowering the rate. 
Unresolved conflicts are preserved in the negotiation trace for 
human review rather than silently absorbed. We argue that surfacing 
more conflicts, even when some require human judgment, provides 
practitioners with a more complete trade-off landscape than resolving 
only a few detected conflicts with high confidence.

\subsection{Implications for Researcher and Practitioner}

For researchers, our results suggest that decomposition 
strategy, whether based on task, knowledge role, or quality dimension, has 
systematic effects on output properties independent of model 
scale, opening an underexplored design space for multi-agent 
RE frameworks. QUARE's protocol shares structural parallels 
with established methods such as Delphi 
(iterative feedback rounds) and the viewpoints 
framework (reconciling distinct 
perspectives), but operationalizes these principles in an 
LLM-based pipeline with explicit conflict categorization. 
A key limitation follows from this analogy: Delphi's 
effectiveness depends on panelist diversity, which QUARE 
approximates only through prompt differentiation over a 
single model. Investigating heterogeneous model configurations 
is a promising direction.

For practitioners, QUARE's quality-specialized agents map 
naturally to organizational review roles (safety officer, 
security analyst, compliance auditor), facilitating integration 
into existing workflows. Its negotiation traces provide an 
auditable record of trade-off decisions, directly supporting 
traceability requirements in safety-critical standards 
(ISO~26262, IEC~62304) and emerging mandates such as the 
EU AI Act (Article~11). The compatibility with lightweight 
models and moderate runtime (55.4s) enables iterative 
development scenarios where requirements are refined 
incrementally.

\section{Threats to Validity}

Aligning with established guidelines for empirical software engineering research, we discuss potential threats to the validity of our study. These threats are organized into four categories: construct validity, internal validity, conclusion validity, and external validity.

\subsection{Construct Validity}

The CHV and MDC metrics rely on LLM-based quality
classification. Blind re-labeling with an independent
classifier yields~75\% agreement and stable metric values,
suggesting robustness. BERTScore is complemented by the
logical consistency score ($S_{\text{logic}}$) to capture
inconsistencies beyond embedding similarity. 

The ISO 29148 quality assessment relies on an LLM-based judge,
which may exhibit systematic biases. We mitigate this through
axis-specific rubrics, independent blind relabeling (75\% agreement),
and a supplementary human evaluation on a stratified sample of 74
requirements, which confirms that the cross-framework ranking is
consistent across both evaluation modes.

\subsection{Internal Validity}
Since neither MARE nor iReDev provides publicly available
implementations, both were reimplemented based on their
published descriptions within the shared OpenReBench
framework under identical conditions. Where the original
specification was ambiguous, we adopted conservative
design decisions. 

Improvements may be driven by increased computational effort rather than architecture. We control model, prompt budget, and environment across all baselines, and isolate contributions via phase-wise ablation. All experiments use three random seeds with averaged results to reduce stochastic bias.

\subsection{External Validity}

The evaluation covers five benchmark systems across
safety-critical, financial, and information-management
domains, though this may not capture the full range of
RE scenarios. The AD case was constructed from public
industrial documentation. As reimplementations may not
perfectly replicate original behavior, results on MARE
and iReDev should be interpreted as reflecting
best-effort reconstruction. Reliance on a single LLM
(\texttt{gpt-4o-mini}) limits generalizability.
\subsection{Conclusion Validity}

Results are averaged over three random seeds (101, 202,
303), with consistent ordering across all five benchmark systems. The absence of formal significance testing
(e.g., Wilcoxon signed-rank test) is a limitation,
constrained by the computational cost of 180 total
LLM runs.

\section{Conclusion}

This paper presented QUARE, a structured multi-agent framework that
decomposes requirements analysis into quality-specialized agents, resolves
cross-quality conflicts through explicit dialogical negotiation, and produces
engineering-ready KAOS goal models with automated compliance verification.
Through a three-way evaluation against MARE and iReDev across 180 experiment
runs, we showed that (1) agent specialization yields up to 2.5$\times$ more
requirements than single-agent generation with balanced quality-axis coverage
(CU\,=\,0.20), while each architectural approach offers distinct diversity
characteristics (RQ1), (2) explicit negotiation achieves the highest semantic preservation (94.9\%) by transforming trade-offs into multi-component solutions that jointly satisfy competing constraints (RQ2), and (3) all frameworks
achieve structural validity, while RAG-augmented verification further yields
98.2\% compliance coverage, a $+$105\% improvement over both baselines (RQ3).

As future work, we plan to apply QUARE to larger-scale industrial settings. In addition, while the current
negotiation protocol draws on dialogical principles, the formal argumentation
literature offers conflict resolution mechanisms such as value-based
argumentation frameworks and abstract argumentation semantics, both largely
unexplored in RE. Grounding the negotiation protocol in these theories could
yield stronger formal guarantees on completeness and admissibility.

\section*{Data Availability Statement}

All experimental artifacts supporting the findings of this study are openly available at: https://anonymous.4open.science/r/OpenRE-Bench-3CF2. The replication package includes framework implementations of QUARE, MARE, and iReDev, prompt templates, benchmark system inputs, raw outputs, evaluation scripts, and human-evaluation data. A structured artifact evaluation guide with reproducibility levels (from output inspection to full 180-run reproduction) is provided in the repository README.

\bibliography{references}
\end{document}